# Transformation media that rotate electromagnetic fields

Huanyang Chen and C. T. Chan

*Department of Physics, Hong Kong University of Science and Technology, Clear Water Bay, Kowloon, Hong Kong, China*

**Abstract:** We suggest a way to manipulate electromagnetic wave by introducing a rotation mapping of coordinates that can be realized by a specific transformation of permittivity and permeability of a shell surrounding an enclosed domain. Inside the enclosed domain, the information from outside will appear as if it comes from a different angle. Numerical simulations were performed to illustrate these properties.

Pendry *et al.* [1] have suggested an interesting idea of using the coordinate transformation approach to design a cloak of invisibility, enabled by mapping the coordinate transformation to electrical permittivity and magnetic permeability which are both spatially varying and anisotropic. The coordinate transformation squeezes space from a volume into a shell surrounding the concealment volume so that the electromagnetic field is excluded from the concealment volume without perturbing the exterior fields. D. Schurig *et al.* [2] verified this idea by calculating the material properties associated with the coordinate transformation and using the results to complete the ray tracing. A full-wave simulation of the circle cylinder cloaking structure with TE polarization was then performed [3] to see these phenomena more visually. A simpler model was experimentally realized later [4]. Recently an optical cloaking device with TM polarization was also proposed [5]. The transformation media concept is drawing much attention as it opens up new possibilities to control the electromagnetic fields [6-8]. Other devices such as the concentrators would be the potential applications of the transformation media as well [9].

In this paper we focus on one kind of 2D transformation media which we would call "rotation coating (RC)". We note that for the cloaking proposed by Pendry, the mapping is from a point to a circle; while for the concentrator, the mapping is from a circle to another circle. The rotation coating performs a rotation of wave fronts and we will define the mapping and explore the physical consequence of this kind of transformation media. We limit ourselves to 2D.

We start from the basic transformation media theory. Suppose that the Jacobian transformation matrix between the transformed coordinate and the original coordinate is [7] [10]:

$$\Lambda^{\alpha'}_{\alpha} = \frac{\partial x^{\alpha'}}{\partial x^{\alpha}}.$$

The associated permittivity and permeability tensors of transformation media become:

$$\varepsilon^{i'j'} = |\det(\Lambda^{i'}_i)|^{-1} \Lambda^{i'}_i \Lambda^{j'}_j \varepsilon^{ij}$$

$$\mu^{i'j'} = |\det(\Lambda^{i'}_i)|^{-1} \Lambda^{i'}_i \Lambda^{j'}_j \mu^{ij}$$

Let us define the following mapping:

For $r < a$, $r' = r$, $z' = z$ and $\theta' = \theta + \theta_0$;

For $r > b$, $r' = r$, $z' = z$ and $\theta' = \theta$;

For $a < r < b$, $r' = r$, $z' = z$ and $\theta' = \theta + \theta_0 \dfrac{f(b) - f(r)}{f(b) - f(a)}$;

which rotates an angle $\theta_0$ for the inner cylinder ($r = a$). The rotational angle is reduced to zero as the radius approaches to $r = b$. Here we assume $\theta_0 > 0$. $f(r)$ could be any continuous function of $r$.

Using the above transformation we find that the permittivity and permeability tensors of the material between $r = a$ and $r = b$ should be:

$$\vec{\varepsilon} = \vec{\mu} = \begin{vmatrix} 1 + 2t\cos\theta\sin\theta + t^2\sin^2\theta & -t^2\cos\theta\sin\theta - t(\cos^2\theta - \sin^2\theta) & 0 \\ -t^2\cos\theta\sin\theta - t(\cos^2\theta - \sin^2\theta) & 1 - 2t\cos\theta\sin\theta + t^2\cos^2\theta & 0 \\ 0 & 0 & 1 \end{vmatrix},$$

where $t = \dfrac{\theta_0 r f'(r)}{f(b) - f(a)} = \dfrac{\theta_0 r}{b - a}$ and $\vec{\varepsilon} = \vec{\mu} = I$ in other places (see Fig. 1). We have dropped the primes for aesthetic reasons.

We could define an ancillary angle $\tau$:

$$\cos\tau = \dfrac{t}{\sqrt{t^2 + 4}}, \quad \sin\tau = \dfrac{2}{\sqrt{t^2 + 4}}.$$

Then, the tensor components can be rewritten as:

$$\varepsilon_{xx} = 1 + 2t\cos\theta\sin\theta + t^2\sin^2\theta = \varepsilon_u \cos^2(\theta + \dfrac{\tau}{2}) + \varepsilon_v \sin^2(\theta + \dfrac{\tau}{2})$$

$$\varepsilon_{xy} = \varepsilon_{yx} = -t^2\cos\theta\sin\theta - t(\cos^2\theta - \sin^2\theta) = (\varepsilon_u - \varepsilon_v)\sin(\theta + \dfrac{\tau}{2})\cos(\theta + \dfrac{\tau}{2})$$

$$\varepsilon_{yy} = 1 - 2t\cos\theta\sin\theta + t^2\cos^2\theta = \varepsilon_u \sin^2(\theta + \dfrac{\tau}{2}) + \varepsilon_v \cos^2(\theta + \dfrac{\tau}{2})$$

with $\vec{\mu} = \vec{\varepsilon}$ completing the material tensor description, where $\varepsilon_u = 1 + \dfrac{1}{2}t^2 - \dfrac{1}{2}t\sqrt{t^2 + 4}$, $\varepsilon_v = 1 + \dfrac{1}{2}t^2 + \dfrac{1}{2}t\sqrt{t^2 + 4}$, $\varepsilon_u \varepsilon_v = 1$.

Here, $\varepsilon_u$ and $\varepsilon_v$ are the principal values of the tensors which are useful for designing this kind of materials. A TE wave experiment would be easier to realize since it is easier to manipulate the permittivity [5]. We note that $\{\varepsilon_u \ \varepsilon_v \ \tau\}$ are functions of t, which in turn depends on r. If we let $f(r) = \ln r$, $t = \dfrac{\theta_0 r f'(r)}{f(b) - f(a)} = \dfrac{\theta_0}{\ln(b/a)}$ is a constant, then $\{\varepsilon_u \ \varepsilon_v \ \tau\}$ all become constants. In our work we will show the TE polarization (we follow the convention in photonic crystal literature, letting $B_z = \mu_z H_z$ with $\mu_z = 1$ [11]) full wave simulation to visualize the useful properties of this kind of transformation media with $f(r) = r$.

We schematically illustrates the geometries in Fig. 1. The incident TE plane wave is from left to right in x direction, whose frequency is 1GHz. The inner radius is $a = 0.25m$ while the outer radius is $b = 0.5m$. The simulated fields shown in the following were computed with more than 250,000 elements and 500,000 unknowns using the commercial finite-element solver COMSOL MULTIPHYSICS.

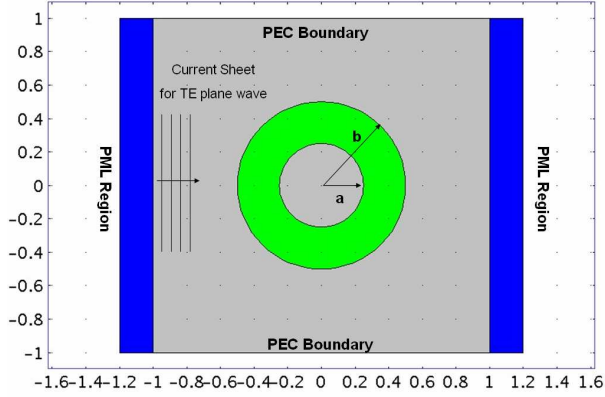

Fig. 1 (Color online) Computational domain and details for the full-wave simulations. Regions in gray are vacuum, regions in blue are PML regions and region in green is the rotation coating. The outside boundaries are PEC boundaries.

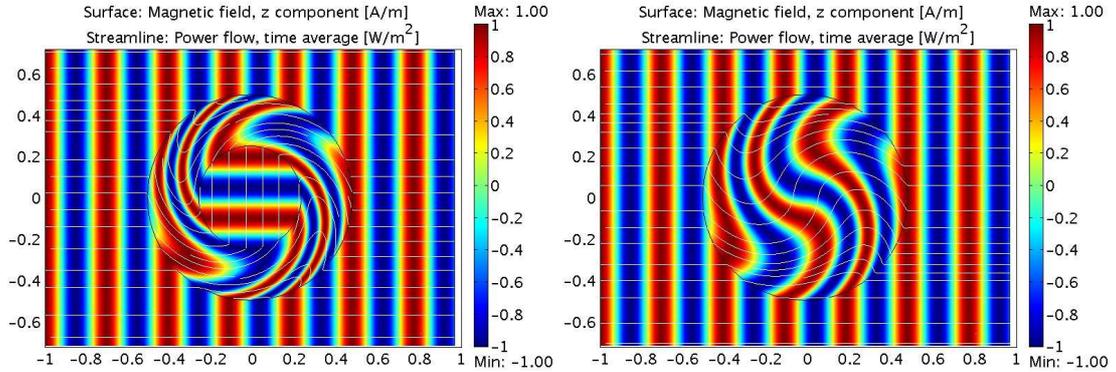

Fig. 2 (Color online) The magnetic-field distribution in the vicinity of the rotation coating. Power-flow lines (in white) show the smooth deviation of EM power. (a)Left: the inner radius $a = 0.25m$; (b)Right: the inner radius $a = 0$.

As an illustration, we set $\theta_0 = \dfrac{\pi}{2}$ as an example. Fig. 2a shows the numerical results for the magnetic-field distribution and electromagnetic power-flow lines. We see that the plane wave changes its direction for $\pi/2$ inside the enclosed domain. If we set $\theta_0 = \pi$, the energy inside and outside the rotation coating flows in opposite directions. It is interesting that we find some turbulence-like pattern in the coating region. Fig. 2b shows the extreme case when we take $a \to 0$. The incoming plane wave splits into two set. Inside the rotation coating, one set has initially a faster phase velocity then slow down, while the other set has initially a slower phase velocity then speed up. If we set $\theta_0$ to be very large, we will find each set of rays going around the origin for many times. In addition, we note that the rotation coating itself is undetectable to a far field observer.

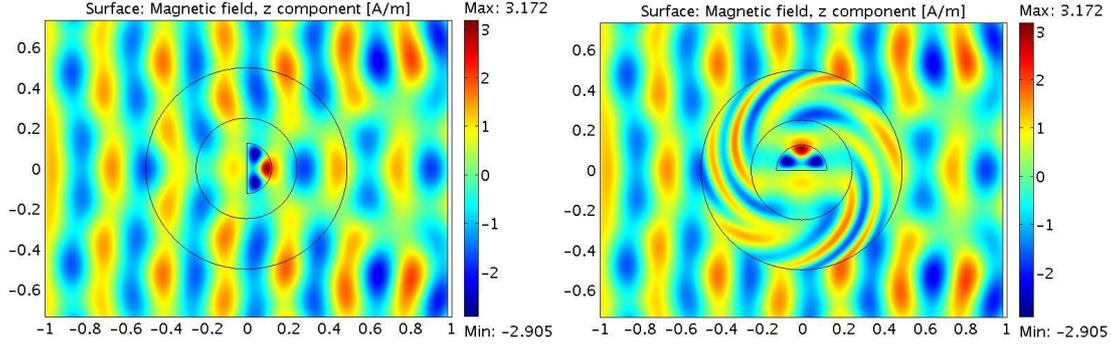

Fig. 3 (Color online) The magnetic-field distribution in the vicinity of the rotation coating. (a)Left: Region in green (see Fig. 3) becomes vacuum, that is scatter without coating outside it; (b)Right: Region in green is still the rotation coating with $\theta_0 = \dfrac{\pi}{2}$, the scatter is rotated from the position in (a) for $\pi/2$ around the origin.

Next, we put a scatterer in the inside domain and see what would be "seen" in the outside world. Firstly, we show a simple scattering problem, with no coating. A small object (a half circular cylinder) with relative permittivity equals to 5 was placed inside and the scattering pattern was shown in Fig.3a. Secondly, the rotation coating with $\theta_0 = \dfrac{\pi}{2}$ was placed outside the object and we rotate the scattering object around the origin for $\pi/2$. The scattering pattern is shown in Fig.3b, and in the far field, the pattern is identical to that shown in Fig.3a. It means when you place an object in the form in Fig. 3b inside the rotation coating, an observer in the outside world would see a rotated image, as the one in Fig. 3a. The image of the scatter inside is rotated $-\pi/2$ from itself around the origin when observed from the outside. Similarly, the image of the scatter outside is rotated $\pi/2$ from itself around the origin when observed from the inside. Observers inside and outside the rotation coating can communicate with each other, but the information is "rotated".

An experimental realization of the rotation coating requires building blocks that have anisotropic dielectric functions. Such materials can in principle be assembled from frequency selective surfaces (FSS) [12]. The similar theory of this kind of rotation mapping could be extended to 3D.

In conclusion, we have shown a new kind of transformation media which rotates the information for a fixed angle, so that observers inside/outside the rotation coating would see a rotated world with respect to each other.

We thank Z.H. Hang, Jeffrey C.W. Lee and Y.R Zhen for useful discussions. This work was supported by Hong Kong RGC through HKUST3/06C. Computation resources are supported by Shun Hing Education and Charity Fund.